\documentclass[twoside,a4paper,10pt]{article}

\usepackage{epsf}

\newcommand{\Sgmm}{\Sigma^0}
\newcommand{\Sgm}{$\Sgmm$\ }

\newcommand{\ASgmm}{\overline{\Sigma}^0}

\newcommand{\Gm}{$\gamma$}
\newcommand{\Lam}{$\Lambda$\ }
\newcommand{\ALamm}{\overline{\Lambda}}

\newcommand{\Ratio}{$\Sgmm/\Lambda$\ }

\newcommand{\etal}{$et$ $al$\ }

\begin{document}

\title{The Ratio \Ratio at RHIC}

\author{G. Van Buren\dag\ for the STAR Collaboration\footnote{For
the full author list and acknowledgements, see Appendix ``Collaborations"
in this volume.} \\
{\it \dag\ Dept. of Physics, Brookhaven National Laboratory,}\\
{\it Upton, NY 11973-5000 USA} \\
{\small E-mail: gene@bnl.gov}}
\date{}

\maketitle

\begin{abstract}
While yields of \Sgm have been measured in
many different colliding systems, no measurements
exist in high energy nuclear collisions~\cite{past}.
It also remains unexplored whether the relative
yields of \Sgm and \Lam change in collisions
of heavy ions where a dense medium may permit
alternate production mechanisms, such as quark
coalescence~\cite{qcoal}. The STAR detector at RHIC is able
to reconstruct the electromagnetic \Sgm decay
into \Lam plus \Gm\ via the weak decay of
the \Lam and \Gm\ conversions into e$^+$e$^-$
pairs in the detector material. We report here on
our measurement of the ratio \Ratio
in minimum bias 200 GeV d+Au collisions.
We also compare the
measured ratio to expectations from
various models.
\end{abstract}

\section{The \Sgm in Nuclear Collisions}

Physics experiments at colliders have mapped production of strange
hadron species over a variety of systems and energies. Such
measurements have commonly included production of the $\Lambda$
but, due to the complexities of its observation, rarely the $\Sgmm$.
These two hyperons share the same quark content, differing only in
isospin and mass, separated by a mere 77 MeV/c$^2$~\cite{PDG}.
Reconstructing the \Sgm requires reconstructing the $\Lambda$, so
comparing the ratio of yields of the two
species\footnote{The yield used here for \Lam will be exclusive of its feed-down
from \Sgm decays, but contributions from resonance decays into
both species will be included.}
provides a simple
test of whether conditions are different for these two similarly-composed hyperons.
Are they produced similarly? Are their final state interactions the same?
Without measurements, these traits may only be assumed.

Above threshold energies for production of $\Sgmm$, the
experimental data for the value of the ratio \Ratio cluster
around 1/3, matching the ratio of isospin degeneracy factors ($1/g_i$) of
the $\Sigma$ and \Lam ground states ($g_{\Sigma}$=3, $g_{\Lambda}$=1),
as will be shown later in this paper.
However there are absences in measurements
for hadronic collisions at moderate
to high energies, and for collisions of sizable nuclei.
High energy nuclear collisions are of particular interest
for the possibility of opening new channels of production
via partonic degrees of freedom~\cite{qcoal,enhance,rate}.

\section{Model Expectations}

Models often start with the production of all particles, followed by the
rapid decay of resonance states. Both the \Lam and \Sgm receive
contributions from decays of strange resonances, but those feeding
the \Lam are much more numerous. Knowledge of
these contributions is pertinent to understanding the full picture,
and efforts to measure these are underway~\cite{Markert}.

\begin{figure}
\begin{center}
\epsfxsize=5.0in
\epsfbox{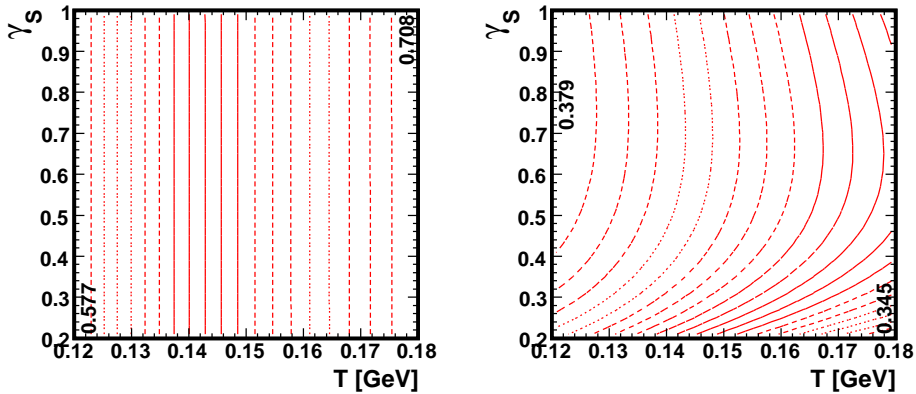}
\vspace{-1.3cm}
\end{center}
\caption{\small \label{fi:RatioCont}\Ratio contours for the primordial (left)
and final (right) ratios as a function of $\gamma_S$ and $T$ in THERMUS.
Minima and maxima of the plots are indicated.}
\vspace{-0.4cm}
\end{figure}

The predictions for \Ratio from a variety of particle
production models are as follows:
the event generator
HIJING/B$\overline{\rm B}$~\cite{HIJING} gives 0.37 for
$\sqrt{s_{\rm NN}} = 200$ GeV d+Au collisions;
the THERMUS~\cite{THERMUS} statistical thermal model gives 0.36
at $T = 160$ MeV, $\gamma_S = 1.0$;
the ``sudden hadronization"~\cite{Letess} thermal model gives $\sim$0.33
at $T = 160$ MeV;
and the ALCOR~\cite{ALCOR} quark coalescence model gives 0.20.
Both thermal models feature little sensitivity to input parameters,
as seen in Figure~\ref{fi:RatioCont} and elsewhere~\cite{Letess},
with nearly no dependence upon chemical potentials.

Though there
is some spread in the predicted values, it should be noted that
the $primordial$ $ratio$ (taken from direct production
before resonance contributions) is much higher. So the final value
of \Ratio may reflect even more upon resonance yields than the
direct production of \Lam and \Sgm. For example,
quark coalescence treats the \Lam and \Sgm equally: the entirety
of their difference then results from the copious resonances~\cite{Levai}.
Thermal models place a penalty on production of more massive states,
so they begin with a primordial value below that of quark coalescence,
but have less resonance contributions to drop the final ratio.

\section{Reconstruction in the STAR Detector}

\begin{figure}
\begin{center}
\epsfxsize=2.4in
\epsfbox{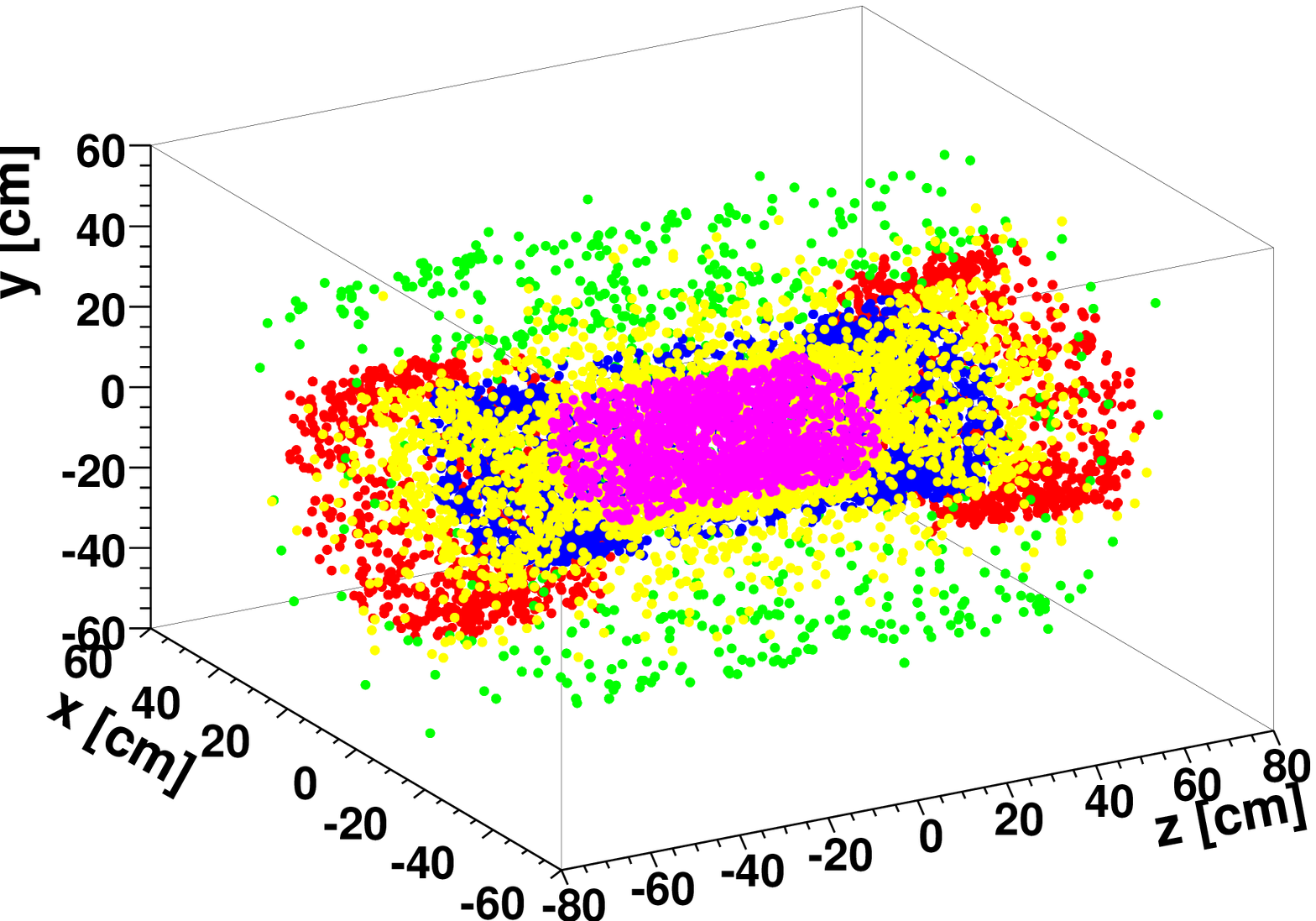}
\epsfxsize=2.3in
\epsfbox{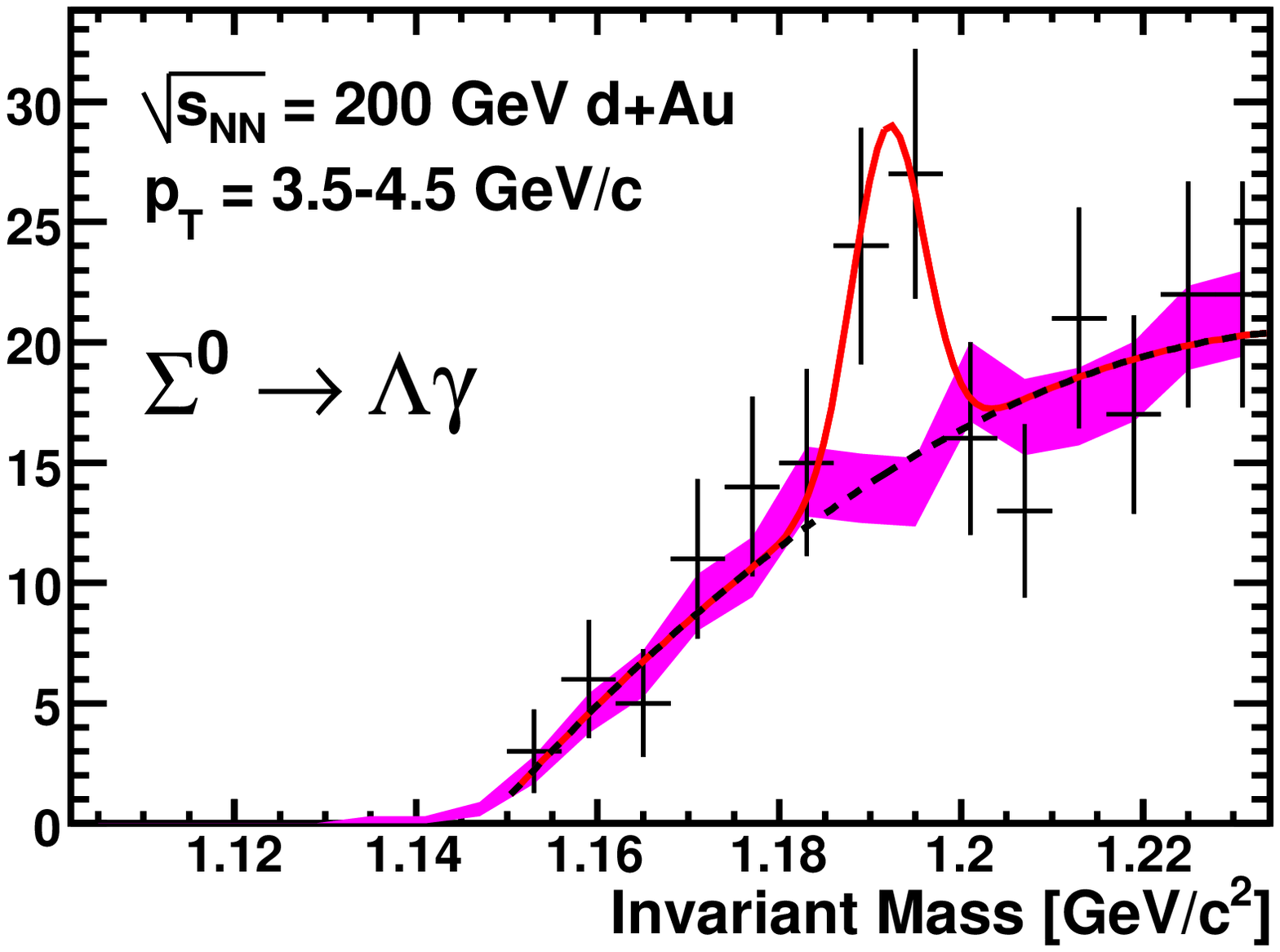}
\vspace{-1.0cm}
\end{center}
\caption{\small \label{fi:SVT}Reconstruction of $\gamma$ conversion
locations reveals inner structure of the STAR Detector (left,
coordinates are in cm about the center of the detector, shadings
represent locations of specific components). \Sgm signal can
be seen via invariant mass of $\Lambda\gamma$ pairs (right,
signal is black crosses, shading is rotated background, lines are
polynomial fit to background plus Gaussian signal.}
\vspace{-0.4cm}
\end{figure}

The STAR Detector is capable of reconstructing the decay
$\Lambda \rightarrow p \pi^-$ with excellent signal-to-noise~\cite{Lam1}.
Reconstruction of photon conversions ($\gamma \rightarrow $\ e$^+$e$^-$)
in detector material can be accomplished similarly~\cite{pi0}, the
validity of which is underscored by the ability to visualize the
structure of detector material from identified conversion points
as seen in Figure~\ref{fi:SVT}(left). Focusing on low transverse momentum ($p_T$)
photons then provides sufficient resolution to identify \Sgm through
invariant mass of $\Lambda \gamma$ pairs when combinatoric
backgrounds are adequately low, shown in
Figure~\ref{fi:SVT}(right). To extract a yield,
the shape of the background under
the invariant mass peak is obtained by rotating candidate \Lam and
\Gm\ daughters from same events to avoid mixing event classes.
Preliminary signals for \Sgm
have been observed via this technique in $\sqrt{s} = 200$ GeV p+p collisions,
as well as $\sqrt{s_{\rm NN}} = 62$ and 200 GeV Cu+Cu
collisions at RHIC, but we concentrate here on
minimum bias 200 GeV d+Au collisions.

\section{Corrections}

By using the same \Lam candidates in both
numerator and denominator of \Ratio, efficiencies for finding the \Lam cancel
out and we are left with three main corrections. The first is weak decay
feed-downs from doubly or triply strange hadrons, which is essentially an
issue for \Lam only. Based on preliminary analyses of $\Lambda$, $\Xi^-$,
and $\Omega^-$ in minimum bias 200 GeV d+Au data in STAR,
approximately 20\% of the \Lam candidates at all $p_T$ come from such
feed-down~\cite{cai}.

\begin{figure}
\vspace{0.5cm}
\begin{center}
\epsfxsize=3.1in
\epsfbox{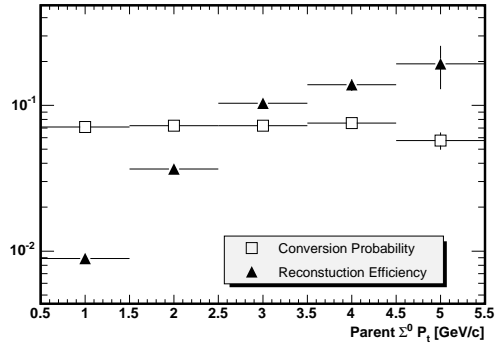}
\vspace{-1.0cm}
\end{center}
\caption{\small \label{fi:effi}Correction factors
for detecting \Sgm photon daughters
as a function of their parent \Sgm $p_T$.}
\vspace{-0.4cm}
\end{figure}

The second and third correction factors are the conversion
probability of photons in detector material and the reconstruction
efficiency of those conversions. These are obtained via embedding
simulated \Sgm decays into real data. The conversion probability
is difficult to determine accurately as it involves simulating every
possible conversion volume properly, but there seems to be
rather little dependence on the $p_T$ of the parent $\Sgmm$.
In contrast, the reconstruction efficiency drops rapidly with
decreasing parent $p_T$, due mostly to reduced acceptance
for low $p_T$ electrons (which are found only for conversions
in the gas volume of the STAR TPC).

To improve statistics, we exploit the expectation that
$\ASgmm/\Sgmm \approx \ALamm/\Lambda$ and sum
baryon+antibaryon counts under the assumption
$\Sgmm/\Lambda \approx (\Sgmm+\ASgmm)/(\Lambda+\ALamm)$.
As the number of reconstructed counts are still small and
bounded by zero yield, the errors are large and asymmetric.
Determination of the measured ratio thus involves a Monte
Carlo allowing values to vary appropriately within their combined
statistical and systematic errors,
and we report the most probable value with errors determined
by the RMS of outcomes above and below it.

\section{Preliminary Results and Conclusions}

\begin{figure}
\begin{center}
\epsfxsize=4.7in
\epsfbox{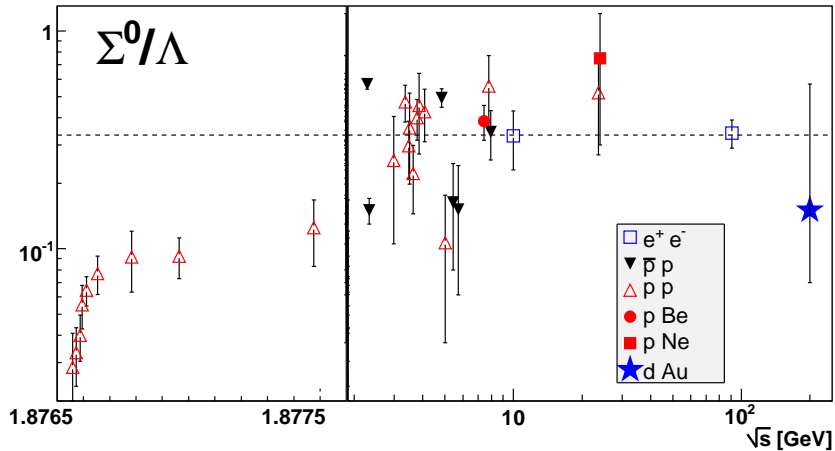}
\vspace{-0.8cm}
\end{center}
\caption{\small \label{fi:PastRes}\Ratio results versus collision $\sqrt{s}$
($\sqrt{s_{\rm NN}}$ for p/d+A)~\cite{past}. Meson-nucleon
reaction results are excluded for clarity, but exist only at intermediate energies
and lie in the same range. The dashed line is
the ratio of isospin degeneracy factors (1/3).}
\vspace{-0.4cm}
\end{figure}

We reconstruct sufficient numbers of \Sgm
to determine \Ratio with significance in two $p_T$ bins
from minimum bias 200 GeV d+Au ($\left| rapidity \right| < 1$).
We find preliminary
\Ratio = $0.16^{+0.57}_{-0.08}$ for \Sgm $p_T$ = 2.5-3.5 GeV/c,
and $0.17^{+0.78}_{-0.10}$ for \Sgm $p_T$ = 3.5-4.5 GeV/c.
Combining the two bins leads to \Ratio = $0.16^{+0.41}_{-0.09}$
and this is shown along with past experiment results
in Figure~\ref{fi:PastRes}. We also find
$\ASgmm/\Sgmm = 0.6\pm0.3$, which does not discount
the summation of baryons+antibaryons for our ratio.
The large error bars on \Ratio prevent definitive model differentiation,
but the low value suggests that strong resonance contributions
are prevalent even in d+Au collisions.

\end{document}